\documentclass[12pt,reqno]{amsart}
\usepackage{amssymb,latexsym,amsmath}
\usepackage{amscd,amsfonts}
\usepackage{epsfig}

\newtheorem{theorem}{Theorem}
\newtheorem{lemma}{Lemma}

\newtheorem{corollary}{Corollary}

\theoremstyle{definition}
\newtheorem{remark}{Remark}
\newtheorem{example}{Example}
\numberwithin{equation}{section}

\begin{document}
\title{Weinhold'length in an isochoric thermodynamic system with constant heat capacity}
\author{Manuel Santoro}\address{Department of Mathematics and
Statistics, Portland State University, PO Box 751, Portland, OR
97207-0751, USA}
 \email{emanus@pdx.edu}
 Dedicated to: Antonino Loiacono
 \qquad Carmela Mazzuti
\begin{abstract}
The purpose of this paper is to study thermodynamic length of an
isochoric two dimensional thermodynamic system with constant heat
capacity. We find that length is related to the heat flow into the
substance. We give examples of Ideal gas and Van der Waals gas.
\end{abstract}\bigskip
 \maketitle
\section{Introduction}
\par
Weinhold proposed a metric$^{12}$ $\eta_{ij}$ in thermodynamic
state space as second derivative of internal energy with respect
to extensive variables $X_{i}$ and $X_{j}$, namely
$\eta_{ij}=\frac{\partial^{2}U}{\partial{X_{i}}\partial{X_{j}}}$
with $i,j=1,...,n$. Such a thermodynamic state space is the energy
surface geometrically described by constitutive relation
$U=U(X_{1},...,X_{n})$ where, for example, $X_{1}=S$,
$X_{2}=V$,etc, with S and V being, respectively, the entropy and
the volume of the thermodynamic system.
\par
Weinhold metric was, then, considered by P. Salamon$^{5,6,7,8}$,
R.S. Berry$^{5,7,8}$, J.Nulton$^{6,8}$, E. Ihrig$^{6}$, and
others$^{7,9}$ to study the physical significance of thermodynamic
length. It was found that the \textit{local} meaning of
$\eta_{ij}$ is the distance$^{7}$ between the energy surface and
the linear space tangent to this surface at some point where
$\eta_{ij}$ is evaluated. Indeed, consider a particular energy
state denoted by coordinates $(X^{0}_{1},...,X^{0}_{n})$. The
tangent space is attached to the energy surface at point
$(U_{0},X^{0}_{1},...,X^{0}_{n})$. Consider, now, a new energy
state $(X_{1},...,X_{n})$ close enough to
$(X^{0}_{1},...,X^{0}_{n})$. Then the distance between the point
on the surface $(U,X_{1},...,X_{n})$ and the tangent space is the
\textit{availability}$^{5,7}$ or the available work$^{5}$ of the
system. This is a local interpretation since it requires just
small displacements, like for fluctuations, from the given point
$(U_{0},X^{0}_{1},...,X^{0}_{n})$ on the surface.
\par
On the other hand, we could consider thermodynamic length taking
the metric $\eta_{ij}$ \textit{globally}. With this approach, we
look at a path $\phi$ on the energy surface between two states
$a_{0}$ and $a_{1}$ and consider the length of the path
\par
\[
L_{a_{0}a_{1}}=\int^{a_{1}}_{a_{0}}[\sum_{i,j}{\eta_{ij}dX_{i}dX_{j}}]^{\frac{1}{2}}
\]
\par
Such a length, in general, represents$^{7}$ the change in mean
molecular velocity depending on the particular nature of the
thermodynamic process defining the path $\phi$ and its dimension
is square root of energy. But thermodynamic length was explicitly
studied$^{7}$ just in the Ideal case. In particular, it was found
that, for a reversible or endoreversible Ideal process at constant
volume, thermodynamic length is given by$^{7}$
\par
\[
L^{v}=2\sqrt{C_{v}}[\sqrt{T_{1}}-\sqrt{T_{0}}]=\sqrt{\frac{\theta}{3}}[\sqrt{v^{*}_{1}}-\sqrt{v^{*}_{0}}]
\]
\par
where $v^{*}_{1}$ and $v^{*}_{0}$ are, respectively, the mean of
the squares of the gas-kinetic velocities at $T_{1}$ and $T_{0}$,
and $\theta=\frac{mC_{v}}{R}$, where m is the particle mass.
\par
In this paper we shall generalize the physical interpretation of
thermodynamic length to a two dimensional thermodynamic system
with constant heat capacity and we shall see that length is
strictly related to the concept of heat flux.
\par
\section{Thermodynamic length with Weinhold metric}
\par
The thermodynamic length between two points $a_{0}$ and $a_{1}$ in
thermodynamic state space is given by the following equation
\par
\begin{equation}
L_{a_{0}a_{1}}=\int^{a_{1}}_{a_{0}}[\sum_{i,j}{\eta_{ij}dX_{i}dX_{j}}]^{\frac{1}{2}}
\end{equation}
\par
where $\eta_{ij}$ are elements of the thermodynamic metric and
$X_{i}$ represent independent coordinates in thermodynamic state
space. We are going to show that thermodynamic length in an
isochoric thermodynamic system with two degrees of freedom with
heat capacity constant is related to the heat flux of a
quasi-static process at constant mole number.
\par
In this manuscript, we shall give, first, a general discussion
about thermodynamic length and, then, we shall focus our attention
to the relation between length and heat flow.
\par
\mathstrut \par Let's consider constitutive relation $u=u(s,v)$
where u is the molar internal energy, s is the molar entropy and v
is the molar volume. s and v are the two independent variables.
Then, it is known that Weinhold metric is given by$^{3}$
\par
\begin{equation}
\eta_{ij}=\frac{1}{c_{v}}
\begin{pmatrix}
 T & -\frac{T\alpha}{k_{T}}\\
 -\frac{T\alpha}{k_{T}} & \frac{c_{p}}{vk_{T}}
\end{pmatrix}
\end{equation}
\par
where
\par
\begin{enumerate}
\item $c_{v}$ is the molar heat capacity at constant volume:
\[
c_{v}=T(\frac{\partial s}{\partial T})_{v}\qquad,
\]
\item $c_{p}$ is the molar heat capacity at constant pressure:
\[
c_{p}=T(\frac{\partial s}{\partial T})_{p}\qquad,
\]
\item $\alpha$ is the thermal coefficient of expansion:
\[
\alpha=\frac{1}{v}(\frac{\partial v}{\partial T})_{p}\qquad,
\]
\item $\kappa_{T}$ is the isothermal compressibility:
\[
\kappa_{T}=-\frac{1}{v}(\frac{\partial v}{\partial p})_{T}\qquad.
\]
\end{enumerate}
\par
Thermodynamic length with such a metric is given by$^{7}$
\par
\begin{equation}
L=\int{[\frac{T}{c_{v}}(ds)^{2}-2\frac{T\alpha}{c_{v}\kappa_{T}}dsdv+\frac{c_{p}}{vc_{v}\kappa_{T}}(dv)^{2}]^{\frac{1}{2}}}
\end{equation}
\par
and, if molar entropy and molar volume are given parametrically as
$s=s(\xi)$, $v=v(\xi)$, then we have$^{7}$
\par
\begin{equation}
L=\int^{\xi_{f}}_{\xi_{i}}[\frac{T}{c_{v}}(\frac{ds}{d\xi})^{2}-2\frac{T\alpha}{c_{v}\kappa_{T}}\frac{ds}{d\xi}\frac{dv}{d\xi}+\frac{c_{p}}{vc_{v}\kappa_{T}}(\frac{dv}{d\xi})^{2}]^\frac{1}{2}d\xi
\end{equation}
\par
Our study is going to focus on length at constant volume which is
given by$^{9}$
\par
\begin{equation}
L^{v}=\int{\sqrt{\frac{T}{c_{v}}}ds}=\int{\sqrt{\eta_{11}}ds}
\end{equation}
\par
\section{Thermodynamic length in an isochoric TD system with constant heat capacity}
\par
\subsection{General discussion}
Consider constant molar heat capacity at constant volume
$c_{v}=T(\frac{\partial s}{\partial T})_{v}$. Then
$(\frac{\partial T}{\partial s})_{v}=\frac{T}{c_{v}}$ and since
$T=(\frac{\partial u}{\partial s})_{v}$, we get the following
equation$^{11}$
\par
\begin{equation}
\frac{\partial ^{2}u}{\partial
s^{2}}-\frac{1}{c_{v}}\frac{\partial u}{\partial s}=0.
\end{equation}
\par
Integrating once we get
\[
\frac{\partial u}{\partial s}=\frac{u}{c_{v}}+f_{2}(v),
\]
with an arbitrary function $f_{2}(v)$. Solving this equation we
get the fundamental constitutive law in the form$^{11}$
\begin{equation}
u(s,v)=f_{1}(v)e^{\frac{s-s_{0}}{c_{v}}}-c_{v}f_{2}(v),
\end{equation}
with another arbitrary function $f_{1}(v)$.
\par
\begin{example}
\textbf{Ideal Gas}
\par
It is known$^{1}$ that for an Ideal Gas
\[
s=s_{0}+c_{v}\ln{(\frac{u}{u_{0}})}+R\ln{(\frac{v}{v_{0}})}
\]
\par
Let $u_{0}=v_{0}=1$ for simplicity. Then, solving for the internal
energy, we get
\par
\begin{equation}
u=v^{-\frac{R}{c_{v}}}e^{\frac{s-s_{0}}{c_{v}}}
\end{equation}
\par
Therefore, considering $(3.2)$, we get the Ideal Gas case if we
set $f_{1}(v)=v^{-\frac{R}{c_{v}}}$ and $f_{2}(v)=0$.$^{11}$
\end{example}
\par
\begin{example}
\textbf{Van der Waals Gas}
\par
The entropy function of the Van der Waals Gas is given by$^{1}$
\par
\[
s=s_{0}+R\ln{[(v-b)(u+\frac{a}{v})^{\frac{c_{v}}{R}}]}
\]
\par
where $a$ and $b$ are positive constants.
\par
Then, solving for u, we get
\par
\begin{equation}
u=(v-b)^{-\frac{R}{c_{v}}}e^{\frac{s-s_{0}}{c_{v}}}-\frac{a}{v}
\end{equation}
\par
Therefore, considering the general case with heat capacity
constant $(3.2)$, we get the Van der Waals Gas case if we set
$f_{1}(v)=(v-b)^{-\frac{R}{c_{v}}}$ and
$f_{2}(v)=\frac{a}{c_{v}v}$.$^{11}$
\end{example}
\par
We have realized, so far, that equation $(3.2)$ is very general as
long as heat capacity is constant. Therefore, our first goal is to
try to express thermodynamic length at constant volume in the same
way.\par Considering $(2.5)$, we have the following relation
\par
\begin{equation}
\frac{\partial^{2}{L^{v}}}{\partial{s^{2}}}=\frac{1}{2c_{v}}\frac{\partial{L^{v}}}{\partial{s}}
\end{equation}
\par
Integrating once we get
\par
\[
\frac{\partial{L^{v}}}{\partial{s}}=\frac{L^{v}}{2c_{v}}+h_{2}(v)
\]
\par
where $h_{2}(v)$ is some arbitrary function of v.
\par
Integrating once more we get the interesting result
\par
\begin{equation}
L^{v}=2[h_{1}(v)e^{\frac{s-s_{0}}{2c_{v}}}-c_{v}h_{2}(v)]
\end{equation}
\par
with $h_{1}(v)$ also arbitrary function.
\par
In this case, $h_{1}$ and $h_{2}$ are completely unknown functions
of molar volume. Nevertheless, we can still recover some
information just about $h_{1}$ in the following way.\par Since
\par
\[
T=(\frac{\partial u}{\partial s})_{v}=\frac{u}{c_{v}}+f_{2}(v)
\]
\par
then, it is easy to see that
$\frac{T}{c_{v}}=\frac{f_{1}(v)}{c^{2}_{v}}e^{\frac{s-s_{0}}{c_{v}}}$
and , therefore,
\begin{equation}
\sqrt{\frac{T}{c_{v}}}=\frac{\sqrt{f_{1}(v)}}{c_{v}}e^{\frac{s-s_{0}}{2c_{v}}}
\end{equation}
\par
This implies that thermodynamic length at constant volume is given
by
\par
\begin{equation}
L^{v}=\frac{\sqrt{f_{1}(v)}}{c_{v}}\int{e^{\frac{s-s_{0}}{2c_{v}}}ds}=2\sqrt{f_{1}(v)}e^{\frac{s-s_{0}}{2c_{v}}}+k(v)
\end{equation}
\par
where $k(v)$ is any function of the molar volume.
\par
This result implies that $h_{1}$ is known for specific examples
being $h_{1}(v)=\sqrt{f_{1}(v)}$, while $h_{2}$ is not. Therefore,
we exactly know what thermodynamic length of a two dimensional
isochoric system at constant heat capacity is up to a constant.
\par
Let's rewrite thermodynamic length in the following way by setting
$k(v)=-2c_{v}h_{2}(v)$,
\par
\begin{equation}
L^{v}=2[\sqrt{f_{1}(v)}e^{\frac{s-s_{0}}{2c_{v}}}-c_{v}h_{2}(v)]
\end{equation}
\par
\begin{remark}
It seems reasonable to think that $h_{2}$ would be somehow related
to $f_{2}$ as $h_{1}$ is to $f_{1}$. But, at this point, such a
statement cannot be made.
\end{remark}
\par
Now, since for an isochoric system
$dq=du=Tds=\frac{f_{1}(v)}{c_{v}}e^{\frac{s-s_{0}}{c_{v}}}ds$,
where $dq$ is the infinitesimal heat or heat flux per unit mole,
then we have the following result
\par
\begin{lemma}
\begin{equation}
(\frac{dL^{v}}{ds})^{2}=\frac{1}{c_{v}}\frac{dq}{ds}
\end{equation}
\end{lemma}
\par
\subsection{Relation between "isochoric" length and heat}
\par
Here we'll derive a relation between the heat flux $q$ as
difference in molar internal energy and thermodynamic length of a
quasi-static isochoric process.\par Since the molar volume is
constant, consider $u_{2}=u(s_{2},v)$ and $u_{1}=u(s_{1},v)$. Let
both $s_{1}$ and $s_{2}$ being greater than the initial entropy
$s_{0}$. It is important to realize that conditions on the molar
entropy $s$ are equivalent to conditions on the temperature since
$T=\frac{f_{1}(v)}{c_{v}}e^{\frac{s-s_{0}}{c_{v}}}$ with both heat
capacity and volume constant. Since $q=\Delta{u}=u_{2}-u_{1}$,
then, considering $(3.2)$, we have
\par
\begin{equation}
q=u_{2}-u_{1}=f_{1}(v)[e^{\frac{s_{2}-s_{0}}{c_{v}}}-e^{\frac{s_{1}-s_{0}}{c_{v}}}]=c_{v}(T_{2}-T_{1})
\end{equation}
\par
with the mass of the substance normalized to $1$.
\par
Similarly, considering the general expression of thermodynamic
length $(3.9)$ we have
\par
\begin{equation}
L^{v}=2\sqrt{f_{1}(v)}[e^{\frac{s_{2}-s_{0}}{2c_{v}}}-e^{\frac{s_{1}-s_{0}}{2c_{v}}}]
\end{equation}
\par
Let's set $s_{1}<s_{2}$ or, equivalently, $T_{1}<T_{2}$. Then,
requiring length to be always positive or equal to zero, we will
consider
\par
\[
1) L^{v}=\int^{s_{2}}_{s_{1}}{\sqrt{\frac{T}{c_{v}}}ds}
\]
\par
\[
2)-L^{v}=\int^{s_{1}}_{s_{2}}{\sqrt{\frac{T}{c_{v}}}ds}
\]
\par
Let's now, for simplicity, set $s_{0}=s_{1}$. Then we can rewrite
both equations $(3.11)$ and $(3.12)$ in the following way
\par
\begin{equation}
q_{in}=f_{1}(v)[e^{\frac{s_{2}-s_{1}}{c_{v}}}-1]>0\qquad
q_{out}=f_{1}(v)[1-e^{\frac{s_{2}-s_{1}}{c_{v}}}]<0
\end{equation}
\par
and
\par
\begin{equation}
L^{v}=2\sqrt{f_{1}(v)}[e^{\frac{s_{2}-s_{1}}{2c_{v}}}-1]\qquad
-L^{v}=2\sqrt{f_{1}(v)}[1-e^{\frac{s_{2}-s_{1}}{2c_{v}}}]
\end{equation}
\par
From these important expression of heat and length it follows,
\par
\begin{theorem}
\par
\begin{equation}
q_{in}=\frac{L^{v}}{4}[L^{v}+4\sqrt{f_{1}(v)}]
\end{equation}
or, equivalently,
\par
\begin{equation}
(L^{v})^{2}+4\sqrt{f_{1}(v)}L^{v}-4q_{in}=0
\end{equation}
\end{theorem}
\par
It easily follows that
\par
\begin{equation}
q_{out}=-q_{in}=\frac{-L^{v}}{4}[L^{v}+4\sqrt{f_{1}(v)}]
\end{equation}
or, equivalently,
\par
\begin{equation}
(L^{v})^{2}+4\sqrt{f_{1}(v)}L^{v}+4q_{out}=0
\end{equation}
\par
We also have the following two corollaries,
\par
\begin{corollary}
\begin{equation}
L^{v}=2[\sqrt{f_{1}(v)+q_{in}}-\sqrt{f_{1}(v)}]=2[\sqrt{u_{2}}-\sqrt{u_{1}}]
\end{equation}
\end{corollary}
\par
which is equivalent to
\par
\begin{equation}
-L^{v}=2[\sqrt{f_{1}(v)}-\sqrt{f_{1}(v)-q_{out}}]=2[\sqrt{u_{1}}-\sqrt{u_{2}}]
\end{equation}
\par
\begin{corollary}
\par
\begin{equation}
 \pm{L^{v}}=0\qquad iff\qquad q=0
\end{equation}
\end{corollary}
\par
Considering equation $(3.11)$, $q=0$ iff $T_{2}=T_{1}$. Therefore,
it does make sense that, at thermal equilibrium, thermodynamic
length $L^{v}=0$.
\par
\section{Examples}
\par
\subsection{Ideal Gas}
\par
In example $1$, we have seen that $f_{1}(v)=v^{-\frac{R}{c_{v}}}$.
Therefore,
\par
\begin{equation}
L^{v}=2[\sqrt{v^{-\frac{R}{c_{v}}}+q_{in}}-\sqrt{v^{-\frac{R}{c_{v}}}}]
\end{equation}
\par
or
\par
\begin{equation}
-L^{v}=2[\sqrt{v^{-\frac{R}{c_{v}}}}-\sqrt{v^{-\frac{R}{c_{v}}}-q_{out}}]
\end{equation}
\par
\subsection{Van der Waals Gas}
\par
In case of Van der Waals gas, we have shown in example $2$ that
$f_{1}(v)=(v-b)^{-\frac{R}{c_{v}}}$. Therefore,
\par
\begin{equation}
L^{v}=2[\sqrt{(v-b)^{-\frac{R}{c_{v}}}+q_{in}}-\sqrt{(v-b)^{-\frac{R}{c_{v}}}}]
\end{equation}
\par
or
\par
\begin{equation}
-L^{v}=2[\sqrt{(v-b)^{-\frac{R}{c_{v}}}}-\sqrt{(v-b)^{-\frac{R}{c_{v}}}-q_{out}}]
\end{equation}
\par
\section{Conclusions}
\par
We have tried to achieve a physical interpretation of
thermodynamic length in a simple isochoric thermodynamic system
with two degree of freedom with constant heat capacity. We also
gave examples like Ideal Gas and Van der Waals Gas.
\par
\section{Comments}
\par
I am aware of the fact that reference [11] has not been submitted
yet. But all the steps needed to the reader for a full
understanding are entirely reported.
\par


\begin{thebibliography}{99}
\bibitem{C} H.B. Callen, \emph{Thermodynamics},
    Whiley, 1960.
\bibitem{DKLR} L. Diosi, K. Kulacsy, B. Lukacs, A. Racz,
\emph{Thermodynamic length, time, speed and optimum path to
minimize entropy production}, arXiv:cond-mat/9503174, 1995.
\bibitem{M2} R. Mrugala, \emph{On equivalence of two metrics in classical
thermodynamics}, Physica A, v.125, 631-639, 1984.
\bibitem{R} G. Ruppeiner, \emph{Thermodynamics: A Riemannian
geometric model} Phys. Rev. A, 20(4), 1608-1613, 1979.
\bibitem{SB} P. Salamon, R.S. Berry, \emph{Thermodynamic Length and Dissipated Availability}, Phys. Rev. Lett., v.51(13), 1127-1130, 1983.
\bibitem{SNI} P. Salamon, J. Nulton, E. Ihrig, \emph{On the relation between entropy and energy versions of thermodynamics length}, J.Chem.Phys., v.80, 436, 1984.
\bibitem{SAGB} P. Salamon, B. Andresen, P.D. Gait, R.S. Berry,
\emph{The significance of Weinhold's length} J.Chem.Phys.,
v.73(2), 1001-1002, 1980.
\bibitem{SNB} P. Salamon, J.D. Nulton, R.S. Berry, \emph{Length in
statistical thermodynamics} J.Chem.Phys., v.82(5), 2433-2436,
1985.
\bibitem{MS} M. Santoro, \emph{Thermodynamic length in a
two-dimensional thermodynamic state space}, J.Chem.Phys., v.121,
n.7, pp.2932-2936., 2004.
\bibitem{MSI} M. Santoro, \emph{On the Helmholtz Potential metric: The Isotherm Length-Work
Theorem}, arXiv,org, math-ph/0404040.
\bibitem{SP} M. Santoro, S. Preston, \emph{Curvature of the Weinhold metric for
thermodynamical systems with 2 degrees of freedom}, manuscript to
be submitted .
\bibitem{W} F. Weinhold, \emph{Metric Geometry of equilibrium
thermodynamics} p.I-V, J.Chem. Phys., v.63, n.6, 2479-2483,
2484-2487, 2488-2495, 2496-2501, 1976, v.65, n.2, pp.559-564,
1976.
\end{thebibliography}
\end{document}